\documentclass[
 ,final            
  ]
  {aipproc}

\layoutstyle{6x9}

\newcommand{\pT}     {$p_{\rm{T}}$}
\newcommand{\pip}    {$\pi^{+}~$}
\newcommand{\pim}    {$\pi^{-}~$}
\newcommand{\pinull}    {$\pi^{0}~$}
\newcommand{\xF}  {$\rm{x}_{\rm{F}}$}
\newcommand{\AN} {$\rm{A_{N}}$}
\newcommand{\Deg} {$^{\rm{o}}$}

\begin{document}

\title{Single Spin Asymmetries in the BRAHMS Experiment}


\author{F.Videb{\ae}k for the BRAHMS collaboration}{
  address={Physics Department, Brookhaven National Laboratory}
}

\begin{abstract}
The BRAHMS experiment at RHIC has the capability to measure the transverese spin asymmetries in
polarized pp induced pion production at RHIC. The first results from a short run show a signaificant
asymmetry for \pip and \pim at moderate \xF. The trend of the data is in agreement with lower energy data 
while the absolute value are surprisingly large.
\end{abstract}

\maketitle

\section{Introduction}

In the last decade or so, measurements of transverse single spin asymmetries in pp collisions 
with polarized beams have attracted much theoretical and experimental interest. 
Results at moderate beam energies\cite{E704} show a sizeable asymmetry up to 
$30\%$ at relative large Feynman-x (\xF) ~and at moderate \pT.
It was expected, naively, from  lowest order QCD estimates that the cross 
sections should have little spin dependence. In order to get a non-zero value both spin-flip amplitudes, 
a phase difference in the intrinsic states as well as a non zero scattering angle is necessary. 
This makes it a higher order effect that can be either in the initial state or in the final state parton scattering.
The asymmetry or analyzing power $\rm{A_{N}}$  is defined as  $(\sigma^{+}-\sigma^{-})/(\sigma^{+}+\sigma^{-})$, 
where $\sigma^{+(-)}$ is a spin dependent cross section for the scattering $pp\to\pi X$, 
and with the spin direction oriented up or down transversely to the beam momentum.scattering plane. 
The target is either un-polarized or the cross sections are averaged over polarization states.
The experiments \cite{E704} has shown that \break $\rm{A_{N}}$(\pip) $>$ $\rm{A_{N}}$(\pinull) $> 0 > $ $\rm{A_{N}}$(\pim).
A recent result from STAR \cite{STAR_An_04} shows a positive $\rm{A_{N}}$ for \pinull at large \xF~ in pp 
collisions at 200 GeV.
The BRAHMS experiment at RHIC is primarily designed and operated to make measurements 
of semi-inclusive spectra of identified hadrons over a wide range in rapidity and \pT.
The PID coverage for pions up to momenta of 40 GeV/$c$ and the option
to measure at 2.3 degrees ($\eta \approx 4$) 
makes it well suited to study Single Spin Asymmetries for identified pions at moderate \xF.
The present contribution presents the first preliminary measurements of $\rm{A_{N}}$  for \pip 
and \pim at moderate values of \xF~ in pp collisions at 200 GeV at RHIC.

\section{Results}

The BRAHMS forward spectrometer consists of 4 dipole magnets, 5 tracking chambers, 
two Time-Of-Flight systems and a Ring Imaging Chrenkov Detector (RICH) for particle identification. 
The angular coverage of the spectrometer is from 2.3 to 15 degrees, and the solid angle 0.8 msr.
Details of experimental setup can be found in \cite{BRAHMS_NIM}.
For transverse spin measurements the kinematic variables of interest are \xF~ and \pT.
Shown in Fig.~1 is the BRAHMS acceptance for the data taken at $\theta = 2.3$ degrees
at the maximum field setting of 7.2 Tm in the spectrometer. 
The momentum resolution $\delta p/p$ is estimated to be $1\%$ at momenta of 22 GeV/c. 
There is an approximate linear correlation between \xF~ and \pT.  It should be pointed out that the acceptance 
does not corresponds to a that of a fixed angle. 
Scattering angles of $2.3^{\rm{o}}$  and $4^o$ are shown on the figure.
Thus care should be taken when comparing to both other experiments (STAR) and to theory.

\begin{figure}[ht]
  \includegraphics[height=.3\textheight]{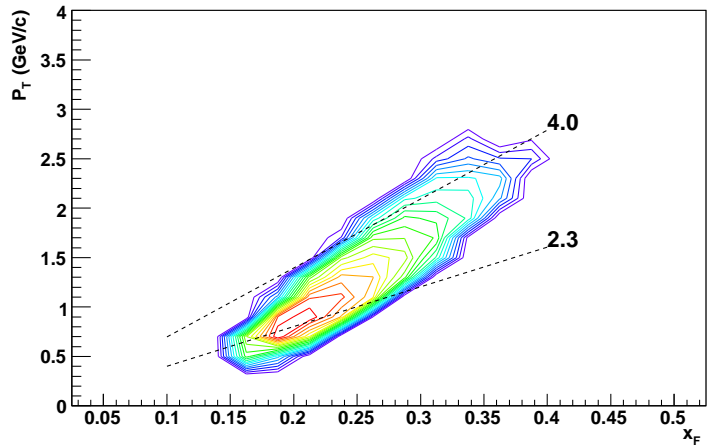}
  \caption{Acceptance in the BRAHMS experiment for pions at the 
nominal setting of 2.3 degrees in \pT~ vs.  ~\xF . The dashed lines indicates the \pT -  \xF~  
correlations for fixed angles of 2.3\Deg~ and 4\Deg, respectively.}

\end{figure}

Tracks were reconstructed from measurements in at least 4 of the 5 chambers, 
and its momentum from 3 independent measurements. 
The tracks are required to project cleanly through the spectrometer.
An approximate vertex can be determined from the timing measurements in sets of symmetricaly placed
scintillator counters (INL) around the beam pipe at 1.5, 4.15 and 6.7 meters \cite{BRAHMS-da_dndeta}. 
The position resolution of the vertex determination is about 10 cm from these measurements. 
In addition vertex positions and live rates are obtained from a set of Cherenkov Counters (BB) with limited acceptance at $\pm 2.15$m
and a pair of Zero Degree Calorimetres (ZDC) placed at $\pm 18$m.  
The tracks accepted in the spectrometer are requiered to point backward to these measurements with an accuracy of 30 cm and
to be within a narrow range of (-40,20) cm of the nominal interaction point.
Due to the measuring angle of 2.3 degree, the spectrometer tends to accepts track weighted 
towards negative vertex positions. 

The particle identification of the pions is done exclusively from the RICH. It is required that the
calculated radius for pion  is within .25 cm from the measured radius, 
and at the same time more than .30 cm away from the estimated radius assuming the track is from a kaon.
This corresponds to about a 2 and 2.5 $\sigma$ cuts, respectively.
Figure 2 shows the radius of rings determined in the RICH for all events, 
and for those identified as pions using the above mentioned cuts in the momentum range $p < $35 GeV/$c$.
The contamination of kaons into the pion sample is estimated to be less than a few percent.
\begin{figure}[h]
  \includegraphics[height=.25\textheight]{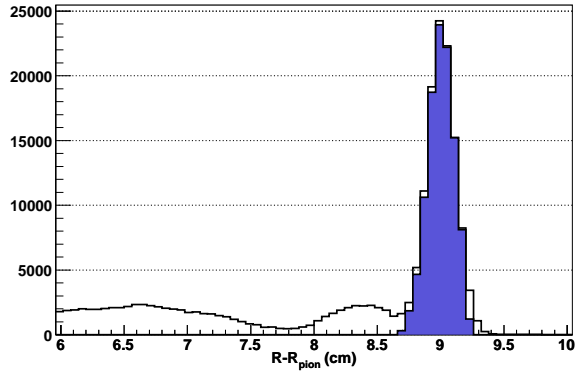}
  \caption{Distribution of radia of the rings in the RICH for accpeted particle in the spectrometer. 
The solid histogram represents those particles identified as pions.}
\end{figure}

In the RHIC accelerator the transverse spin polarization is altered between the
56 bunches of polarized protons that forms the beam in each of the two rings. 
Thus most experimental time-dependent effects originating from the spectrometer and the
vertex determination cancel out when constructing the raw asymmetries
 $$\epsilon =(N^{+}-L*N^{-})/(N^{+}+L*N^{-})$$ The $N^{+(-)}$ represents the yield of pions in a given 
kinematic bin where the beam spin direction is up or down relative to the reaction plane determined by
$k_{beam}\times k_{out}$. 
The factor $L$ is the ratio of the luminosity of bunches with positive polaraization to those of negative polarization thus accounting for  
 non-uniform bunch intensities.
The luminosity ratio is determined independently from the spectrometer data using several
measures of collision rates from the INEL, BB, and ZDC detector systems. 
It is estimated that the systematic error from the relative luminosity 
measurements is in order $0.5\%$.
\begin{figure}[h]
  \includegraphics[height=.3\textheight]{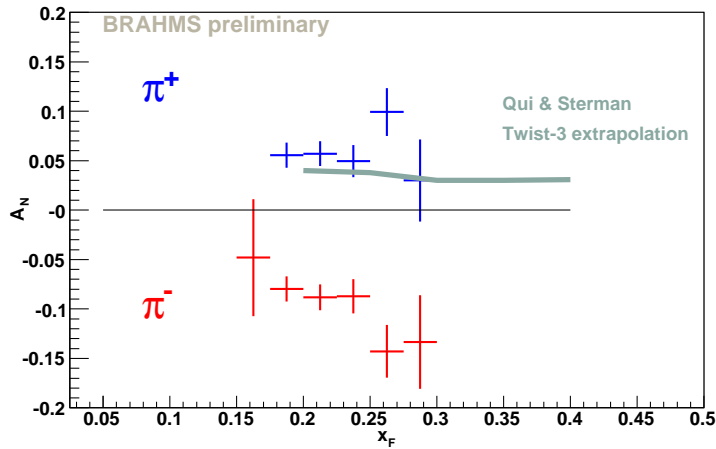}
  \caption{Analyzing power \AN~ for \pip and \pim.}
\end{figure}

The asymmetry is the in turn determined from \AN$=\epsilon/P$. 
The polarization (P) as determined from the CNI measurements \cite{CNI} is $\approx 42\%$  
for the \pip measurements and $\approx38\%$  for the for \pim measurements in the stores used. 
The systematic error on beam polarizations is $\approx  15\%$ and represents a scaling error on the values of
\AN. 
This error is expected to be reduced after the final analysis of CNI data. 
The measured raw asymmetries corrected for the beam polarization is shown in Fig.~3 for \pim and \pip.
The \pip asymmetries are positive while the \pim are negative i.e.  the same sign dependence
 as seen in the E704 data at lower energy. 

Several theorectical models have been worked out for the single spin asymmetries to clarify the
importance of inital vs. final state effects as put forward through the Sivers and Collins effects.
 In this contribution we compare the data vs. extrapolations of twist 3 (initial state) calculations by Qiu 
and Sterman \cite{QuiSterman}. The pQCD calculations are apriori not valid at the lower values of \pT~ covered
 in the present measurements. Never the less it gives a good estimate how kinematic cuts may effect
predictions as to give rise to a near constant \AN~  in a limited range of \xF. 
Both the magnitude and the \xF ~dependence is in reasonable agreement with the data.

During run-5 RHIC has delivered much increased integrated luminosity and with a larger 
beam polarization  of $45-55\%$. BRAHMS has added new vertex detector that will provide a global
vertex resolution of $\approx 2$ cm. 
Data have been recorded for \pip and \pim in the \xF~ range of 0.15-0.35 with about 10-20 times 
the statistics in the data presented here.

In summary, the first data from polarized proton data from BRAHMS were obtained in the RHIC Run-4 and shows a finite \AN~ 
for \pip and \pim with sign ordering as observed previously in E704 at FNAL. 
In addition the protons are found to have \AN $ \approx 0$.
Data from the ongoing RHIC Run-5 will give an order of magnitude better 
statistics and should enable BRAHMS to make comparisons to several theoretical models

This work is suppord by the Division of Nuclear Physics of the Office of Science of 
the U.S. Department of energy under contract DE-AC02-98-CH10886,
 the Danish Natural Science Research Council,
the  Research Council of Norway, the Jagiellonian University Grants 
and the Romanian Ministry of Education and Research. I will like to thank J.Qui for supplying 
the calculations for the twist-3 extrapolations, and Brendan Fox for advice, calculations 
and help with the experimental setup for the first attempt to collect spin sorted data with BRAHMS spectrometer 
in 2003, a setup that was used in the brief 2004 run that resulted in the present data.

\end{document}